\documentclass[12pt,pra,aps,amssymb,amsfonts,amsmath,tightenlines]{revtex4}

\usepackage[caption = false]{subfig}
\usepackage{graphicx}
\usepackage{dsfont} 
\usepackage{bbm} 

\newcommand{\rd}{\mbox{$\rm d$}}

\usepackage{xcolor}
\usepackage{bm}
\usepackage{txfonts}
\usepackage{amsfonts}
\usepackage{braket}

\begin{document}

\title{Biological efficiency in processing information} 
\author{Dorje C. Brody$^{1}$ and Anthony J. Trewavas$^{2}$}
\affiliation{
$^1$School of Mathematics and Physics, University of Surrey, 
Guildford GU2 7XH, UK} 
\affiliation{
\vspace{-0.3cm}
$^2$Institute of Molecular Plant Science, University of Edinburgh, 
Edinburgh, UK
}

\vspace{0.2cm} 
\date{\today}

\begin{abstract}
\noindent 
Signal transduction, or signal-processing capability, is a fundamental property of nature that manifests universally across systems of different scales -- from quantum behaviour to the biological. This includes the detection of environmental cues, particularly relevant to behaviours of both quantum systems and green plants, where there is neither an agent purposely transmitting the signal nor a purposefully built communication channel. To characterise the dynamical behaviours of such systems driven by signal detection followed by transduction, and thus to predict future statistics, it suffices to model the flow of information. This, in turn, provides estimates for the quantity of information processed by the system. The efficiency of biological computation can then be inferred by measuring energy consumption and subsequent heat production.  
\end{abstract}

\maketitle

\section*{Introduction} 

Plants in wild circumstances commonly experience a fluctuating environment 
that can threaten the survival of any individual and thus limit its capability to 
reproduce. Within reasonable environmental limits, plants react to such 
variations by changing their behaviour adaptively, improving their probability 
of survival \cite{Dobzhansky}. They detect and transduce the information they 
receive from their environment and use that processed information to 
establish counter-reactions in behaviour to improve survival. Unlike animals 
that can move out of danger, plants in general cannot move away from 
extreme threats except 
via distribution of seed or pollen or the production of dormant underground 
organs such as tubers, fleshy roots and stolons that weather the hazard and 
produce new shoots when conditions change. When milder threats are 
experienced from numerous biological and abiological threats, developing 
plants counteract by changing development (phenotypic plasticity) or 
molecular physiology to improve the probability of survival \cite{Gilroy}. In 
1871, Charles Darwin \cite{Dawin} commented that ``intelligence is based 
on how efficient a species becomes at doing the things they need to survive,'' 
leading to the general recognition that ``adaptive behaviour is the essence of (biological) intelligence'' \cite{Anastasi,Beer}. Plants use the information that 
they receive from their noisy environment to construct adaptive responses. 
Those species that do so more efficiently may benefit in the struggle for 
existence. But in establishing the reality of Darwin’s view, there is a 
necessity to quantify the environmental information that any plant 
experiences and in turn be able to measure how efficiently it is processed. 
It is obvious that all plants experience information from their environment 
and that is commonly stated, but without attempts to quantify what is 
experienced, the term really has little meaning. Many areas of plant 
science would benefit if this can be achieved as would understanding 
of biological intelligence. 

Intensive research in the last 30 years has identified the central role of 
cytosolic calcium $[{\rm Ca}^{2+}]_{\rm i}$ in transducing most 
environmental signals experienced by plants \cite{Gilroy,Kudla}. Signals 
are perceived by specific receptors often located in the plasma membrane. 
Occupied receptors, it is assumed, form membrane patches that initiate 
intracellular calcium transients, $[{\rm Ca}^{2+}]_{\rm i}$, within less than a second and lasting at the very most a minute. Inhibition of these 
transients prevents the subsequent physiological or phenotypic changes 
that normally occur hours, or even days or weeks later. 
These transients can oscillate if the environmental change is prolonged 
\cite{Gilroy,Kudla,Delormel}. 
Cytoplasmic changes in $[{\rm Ca}^{2+}]_{\rm i}$ are perceived by well 
over 100 proteins (CDPKs, calcium-activated kinases, calcium-calmodulin 
kinases, calmodulin, calmodulin-like proteins, CBL kinases, 
calcineurin-like proteins) that bind ${\rm Ca}^{2+}$ directly. A variety of 
channels, pumps, exchangers and carriers are responsible for 
${\rm Ca}^{2+}$ entry. These membrane-bound proteins help shape and 
change the internal spatial distribution generating a distinctive signature 
often characteristic of the specific signal. It is thought that different signals 
recruit different combinations of these ${\rm Ca}^{2+}$ binding proteins 
that continue information transfer into hundreds of protein kinases of 
differing kinds that manipulate further downstream change into altered 
ion fluxes, enzyme activities, cytoskeletal reconstructions, and ultimately 
particular combinations of gene expressions.

Whatever the mechanism by which information is detected, transmitted, 
processed, stored, and in some cases erased, the amount of information 
processed can be estimated from dynamical behaviours resulting from 
the processing of information, and the efficiency of biological computation 
can be inferred from the amount of energy consumed by the cells to 
process information. That is, these estimates can be made without 
necessarily the knowledge of detailed structures and mechanisms for 
achieving the detection of environmental conditions by plants \cite{Diaz}. 
The purpose of the present paper is to explore this line of investigation.

\section*{Information in wild environments will be noisy} 

From the point of view of mathematical modelling of adaptive behaviour, 
what underpins adaptation is the detection and processing of noisy 
information about environmental conditions. That is, adaptation to changes 
of external conditions is not possible without the information about those 
noisy conditions. The processing of information, in turn, leads to adaptive 
dynamics, because the state of a biological system can be viewed as the 
reflection of the system’s perception of the condition of its environment. 
Thus, we are led to the hypothesis that dynamical behaviours of all 
biological systems originate in signal detection, and that by modelling 
the noisy information flow for environmental cues, it is possible to predict 
the statistics of biological dynamics by means of a logical deduction 
based on optimal (efficient) signal detection. 

The information-based approach to modelling biological dynamics allows 
for a way of measuring information contents of environmental cues, or 
messages, that biological systems such as green plants experience. 
When environmental conditions change, much of the processed information 
about previous conditions will presumably be erased. For instance, when a 
climbing plant detects an object that can support its growth, but then the 
object is removed, it seems likely that this information is erased. Otherwise, 
plant cells will require considerable memory capacity to store all the 
day-to-day processed information, which seems unlikely. Of course, if 
erasure of processed information results in energetic penalties, for 
instance the leaf-folding behaviour of \textit{Mimosa pudica} under false 
alarms, then there is an advantage in storing processed information, but 
even here the memory appears to be erased within thirty days 
\cite{GRDM}. However, if such processed information 
is regularly erased, then as a consequence we can estimate the 
computational efficiencies of biological systems. This is  because 
information erasure requires energy consumption indicated by additional 
heat production, and this can be calculated and potentially measured, 
and compared to the total energy consumption and heat production of 
biological cells. 

\section*{Such adaptive dynamics also has relevance at the quantum level} 

Perhaps unexpectedly, this adaptive dynamics appears to have a deeper, quantum-mechanical origin. Consider a quantum system, like a particle, in an external field, say, a photon field. For the state (wave function) of the system to evolve in time in accordance with the conditions of the external field, it is required to extract information about the condition of the environment. This information is necessarily noisy, but if we assume that nature is intrinsically efficient in the sense that dynamical evolution is dictated by the criterion of minimising the degree of uncertainties about the condition of the environment, then the system must follow the dynamics that is consistent with optimal signal detection. Such a picture for the dynamics of a quantum system is not entirely dissimilar to, say, a plant cell extracting the information of the light source from noisy interaction with a photon field for the phototropic or heliotropic dynamics. Indeed, there is a great deal of parallelism between quantum and biological \textit{dynamics} (not to be confused with static aspects such as quantum coherence or entanglement that may or may not be of importance in biology), and this can be exploited to pave the way towards a better understanding of information and signal transduction in biology, because the theory of quantum dynamics in noisy environments is well developed, while biological dynamics is less understood.  

But how is such a characterisation of quantum dynamics as extracting information from the environments compatible with the more traditional formulation of quantum systems by means of Schrödinger’s wave equation? And can we show that nature is indeed efficient in its ability to process information? If so, what are the implications in biology, if any, and what do we learn from these observations? 

The purpose here is to clarify some of these issues, with the view towards 
building a unified information-based approach to modelling quantum and biological systems, and in particular, plant dynamics. Specifically, we begin by explaining how the specification of the information flow can be used to model dynamic features of biological systems. In particular, focusing on the response of green plants to environmental cues, it is shown how the techniques of information and communication theory can be used to characterise their dynamics, even though in the context of signals from environments there are no agents purposely transmitting these signals. The sun does not emit photons with the purpose of letting green plants on earth know of its whereabouts.  Wind does not blow at high speeds for the purpose of stressing plant structures. The environments, in effect, carry the signals, but they are not purposefully built like a telephone line carrying signals. 

We then show that precisely the same mechanism of signal-detection-induced dynamics is already present at the quantum-mechanical level. After illustrating how the orthodox views of quantum theory requires a closer re-examination, taking into account the impact of environments -- more precisely, of noise -- 
we show how the modified formulation of quantum theory is underpinned by optimal signal detection problems. In particular, it is shown how signal detection leads to the so-called decoherence effects of quantum mechanics. We then discuss the advantages of applying models describing dynamics of quantum systems for the characterisation of biological dynamics, when there is an incompatibility between adaptation and the rate of change in the environment. Such dynamical models not only allow for the prediction of future statistics but also enables the quantification of information contents that are otherwise difficult to estimate from experimental data. We conclude with some discussion on open challenges in biology from an information-based perspective. 

\section*{Processing of noisy information induces dynamics in biology} 

Information theory is traditionally concerned with the amount of information 
contained in a signal transmitted through a noisy channel, the informational 
capacity of the channel, and corrections of errors \cite{Kullback,Cover}. 
These ideas 
refer to an established communication channel used by a sender and a 
receiver of the message, which, in biological contexts, might translate into 
signal transduction in molecular and cellular networks \cite{Cheong}. 

As well as internal communication channels, biological systems constantly 
extract information from their environments, in the absence of a transmitter 
of a message or a communication channel designed to deliver specific 
environmental cues \cite{Mescher}. Take, for instance, the case of green 
plants. Whether it is sunlight, scattered photons, volatiles, wind, water, 
soil nutrients, or gravity, information about these environmental cues is 
processed effectively \cite{Trewavas}. Even when there are well-defined 
communication channels, estimating the amount of information transmitted 
from data is not straightforward \cite{Thomas}, so the problem becomes 
even less tangible when it comes to quantifying information processed by 
plants through what seems to be ambiguous communication channels. 
This is where a mathematical model can be exploited to better understand 
the link between information processing and resulting biological dynamics. 

Processing of information commonly leads to phenotypic changes and movements in plants. For instance, leaf blade reorientation towards the 
direction of the sun. Such a dynamical behaviour is induced by detection 
and processing of signals encoded in photons. Now in the conventional 
approach to mathematical modelling of biological dynamics, one typically 
attempts to model the resulting motion directly in the form of differential 
equations, which in some cases may be superimposed with additional 
noise terms. While such an approach of directly modelling observed 
dynamics has seen some successes in modelling certain specific systems, 
for, models can always overfit to match simple dynamics, ultimately it 
appears futile to attempt to directly model dynamical behaviours 
of biological systems driven by processing of noisy information. 

The reason is a simple one. What underlies biological dynamics is the noisy 
information about environmental conditions, which are processed to arrive at 
choices, and hence dynamics. The input, therefore, is the specification of 
noisy information, on which optimal signal detection can be applied to arrive 
at the output, and this leads to phenotypic change and/or movement. This 
input-output transformation is typically highly nonlinear. Hence any attempt 
to model the output will fail to capture the input-output causal relation. In 
other words, direct modelling of the dynamics in most cases cannot replicate 
the underlying information-processing features.  

The way forward, therefore, is to model the noisy flow of information, from 
which the implied dynamical behaviour of a given biological system can 
be \textit{deduced} (rather than modelled) using the theory of optimal 
signal detection. In this way the causal structure underlying the resulting 
biological dynamics becomes transparent. 

\section*{Environment acting as a communication channel} 

Of course, in many cases there are no purposefully built 
communication channels when it comes to biological systems processing 
information about environmental cues, and this fact has perhaps prevented 
the application of communication theory to describe biological dynamics. To 
proceed, therefore, we must regard the environment 
as a whole, \textit{in effect}, as playing the role of a communication channel. 
The idea is as 
follows. Take the case of a green plant. There is a range of quantities of 
interest that plants wish to identify, such as the location of the light source, 
the direction of the gravitational pull, the gradient of volatile concentration 
in the air or soil, the existence of plants or other objects in their vicinities, 
and so on \cite{Trewavas2}. 
We may refer to these quantities as `signals' or `messages', even though in 
none of these cases, an active agent is transmitting the signal. Yet, 
these signals do exist, not in abstraction but in reality, and they are carried 
by the environment (air molecules, photons, gravitons, and so on) to reach, 
along with noise (e.g., scattered photons), 
the plant receptors. 

From the viewpoint of communication theory, 
whether there are active transmitters of signals or purposefully built channels 
is immaterial, because information has been successfully transmitted, albeit 
superimposed with noise. It is the environment itself that generates what 
plants perceive as signals, and it is the environment itself that carries these 
signals along with noise.  

\section*{Dynamics following from Bayesian updating} 

In many 
cases, the signal (i.e. the quantity of interest from the plant's perspective) 
is given by a time series, and so is noise, but for an illustrative purpose take 
for simplicity the case 
in which they are fixed in time. In more specific terms, let $X$ be the random 
variable representing the signal, and suppose that it can take $N$ discrete values 
$\{x_i\}$ with the \textit{a priori} probabilities $\{p_i\}$ (the continuous case 
can equally be treated). Similarly, let $\epsilon$ represent noise, with the 
density function $f(x)$. Information acquisition in signal detection is then 
modelled by learning the value $\xi=X+\epsilon$ of signal plus noise. The 
motion of plant (or any other biological system for that matter) resulting from 
information acquisition will then be consistent with the transformation from 
the \textit{a priori} mean of $X$ to its \textit{a posteriori} mean \cite{Wiener}: 
\[
\sum_{i=1}^N x_i p_i ~\Rightarrow~ 
\sum_{i=1}^N x_i  \left( \frac{p_i \, f(\xi-x_i)}{\sum_{j=1}^N p_j \, f(\xi-x_j)} 
\right). 
\] 
Here, the term given in the bracket on the right side is just the Bayes formula 
for the \textit{a posteriori} probability of $X$, subject to the specification of 
partial information $\xi$ about $X$. 

To gain a better intuitive understanding of the information-based 
formulation, consider the 
experiment on the parasitic plant \textit{Cuscuta pentagona} (dodder) to 
demonstrate 
its ability to detect the host location via inferring the gradient of volatile 
chemical concentration in the air \cite{DeMoraes}. The initial motion of the seedling as circumnutation is dictated by the genetic information already encoded in the plant (as opposed to processed information), and suggests that the 
\textit{a priori} distribution of the host location is uniformly distributed over a circle centred at the point of germination, on account of the symmetry of the configuration. 
But then the detection of 
volatiles in the course of nutation turns the \textit{a priori} distribution into 
the \textit{a posteriori} distribution, resulting in the observed directed growth 
orientation towards the host location \cite{DeMoraes}. The \textit{a posteriori} 
distribution can be estimated accurately by increasing the sample size 
considered in \cite{DeMoraes}. The amount of information processed can 
then be identified, in units of bits, by working out the relative entropy or 
entropy reduction associated with the \textit{a priori} and the \textit{a posteriori} 
distributions. 

This example illustrates how a seemingly abstract mathematical idea of 
signal detection can be used to characterise the dynamical behaviours of 
biological systems. This quintessentially biological feature of `intelligently' 
adapting to environmental conditions, in the sense of optimally processing 
noisy information so as to reduce, on average, the degree of uncertainty as 
measured by entropy, however, seems to have a deeper 
quantum-mechanical origin. That is, dynamical features of systems induced 
by signal transduction appears to be universal across all scales. 

\section*{Role of environmental noise in quantum dynamics} 

In physical science the dynamics of a system is determined by the 
specification of its energy. That is, once the system energy is specified, 
then equations of motion for the state of the system unambiguously 
determine how the initial state evolves in time. 
From Newtonian mechanics to quantum theory, this approach has 
proven incredibly effective. Quantum theory, in particular, is one of the most 
stringently tested theories in science, enabling high-accuracy predictions for 
experimental data. Yet, there are reasons to doubt whether this energy-based 
description of physical systems, supplemented with an absolute 
determinism, is complete in the quantum realm. What lies 
at the heart of the issue is an internal contradiction in the conventional 
characterisation of 
a quantum system. Namely, a quantum system in isolation is defined to be one 
that is modelled by its Hamiltonian (energy), which determines the system 
dynamics according to the Schr\"odinger equation. In most example systems, 
however, their Hamiltonians represent system-environment interactions, 
such as the effects of external fields; but this contradicts the assumption that 
the system is in isolation.  

From the perspective of a quantum system, the interaction with its 
environment, more specifically, is interaction with environmental particles 
such as photon fields, and this is inadvertently noisy owing to the existence 
of the very large number of particles involved. 
The omission of the influence of noise in the theoretical description (such as 
Schr\"odinger's theory) in turn leads to one of the central 
issues in contemporary physics of explaining the transition from quantum to 
classical, as exemplified by the so-called measurement problem of quantum 
theory \cite{Isham}. 
For sure if a system is in equilibrium with its environment, then it is 
expected that the system energy is conserved \textit{on average} subject to 
fluctuations due to noise, and that the degree of fluctuation scales down in 
the limit whereby the energy gaps of the system are sufficiently small 
(cf. \cite{Adler}). This leads to the 
hypothesis that one should regard Schr\"odinger's theory as an 
approximation to a more realistic description of the quantum world in which 
the role of noise is taken on board. 

Of course, any deviation from the noise-free deterministic `unitary' dynamics 
(if the relative separation or difference of two wave functions remains 
constant in time, then the dynamics is called unitary) 
predicted by Schr\"odinger's theory must be subject to scrutiny. 
This is because if the Schr\"odinger equation for 
the wave function of the system is replaced with a stochastic dynamical 
equation that reflects the impact of noise, then experimental predictions 
will in general deviate away. However, for a number of prototypical 
quantum systems for which ample experimental data exists, this deviation 
has been shown to fall within error bars, and hence not detectable, provided 
that the random dynamics of the wave function conserves 
energy expectation \cite{Adler}. 
In other words, there is a family of stochastic models 
describing the dynamics of quantum systems that  
does not contradict the successful predictions arising from the 
traditional quantum theory, while incorporating the impact of noise. 

This may seem surprising, but the intuitive reason is that 
the timescale for such a stochastic 
dynamics to deviate significantly away 
from unitarity is inverse proportional to system's energy variance \cite{H}. 
Hence 
for a small quantum system, for which the energy variance is small, the 
stochastic evolution will maintain approximate 
quantum coherence for long times. On the other hand, 
if the system is coupled to a measurement apparatus, then the energy 
variance is 
amplified by the uncertainties of the macroscopic pointer state 
(the state of the pointer of the apparatus pointing in one direction or another 
will result in huge energy uncertainty), forcing 
the wave function to `collapse' into one of the stable eigenstates instantly. 

\section*{Characteristics of quantum dynamics induced by signal detection} 

What then is the main trait of such stochastic models? 
It can be shown that for the class of stochastic 
Schr\"odinger equations for which the energy expectation is conserved, 
their solutions can be obtained by formulating underlying signal detection 
problems about the energy of the system \cite{BH1,BH2}. 
Suppose that a quantum system (say, a particle) is 
immersed in an external field, and the interaction of the two is modelled 
by a Hamiltonian ${\hat H}$, which characterises the energy of the system. 
One can think of ${\hat H}$ as a matrix with 
eigenvalues $\{E_i\}$, representing definite energy values. An energetically 
definite (certain) state of the system is represented by the eigenvector 
$|E_i\rangle$ of ${\hat H}$ corresponding to the eigenvalue $E_i$. A generic 
state $|\Psi\rangle$ of the system can be expanded in terms of the 
eigenvectors of the Hamiltonian: 
\[ 
|\Psi\rangle = \sum_{i} \sqrt{p_i}\, |E_i\rangle . 
\]
Here, we have ignored the phase factors that are quantum-mechanically 
important but not relevant to the ensuing discussion. The expansion 
coefficients $\{\!\sqrt{p_i}\,\}$ 
are such that their squares add up to one. 
These are then interpreted as determining the probability of finding the 
system to be in the state with a definite energy. Such a generic state is 
therefore a state of indefinite energy. Our hypothesis then is that through 
interaction with the 
environment, however, the system acquires partial information about 
energy -- partial because of the prevalence of noise.  

Typically the information acquisition occurs continuously in time, but for 
simplicity consider a `single-shot' transfer of information. The (unknown) 
quantity of interest is the energy, which we model by a random variable 
$H$ taking the value $E_i$ with the probability $p_i$. We let $\epsilon$ 
model the noise, which is assumed to be independent of the `signal' $H$ 
and has the density $f(x)$. Then information transfer from the environment 
to the system can be modelled 
by the specification of the value $\xi=H+\epsilon$ of the sum of signal and 
noise. There are two unknowns, $H$ and $\epsilon$, and one known, $\xi$, 
so this data is insufficient to determine the value of $H$, but it can be used 
to reduce the uncertainty of $H$, just as in the biological case. 
Such a reduction results in updating the prior probabilities $\{p_i\}$ into 
posterior probabilities $\{\pi_i\}$. If this updating of information is efficient in 
that it maximally reduces the uncertainty, then $\pi_i$ must be the probability 
that $H=E_i$ conditional on the value of $\xi$, and this can be worked out by 
use of the above-mentioned Bayes 
formula. Hence the state of the system, after acquiring this information, and if 
nature is efficient, is given by 
\[ 
|\Psi'\rangle = \sum_{i} \sqrt{\pi_i}\, |E_i\rangle,
\]  
where the conditional probability 
$\pi_i=\pi_i(\xi)$ is a function of the information transmitted. The 
acquisition of information by the system, represented by the transformation 
$p_i\to\pi_i$, then results in the well-known decoherence effect of quantum 
theory, as described below. 

\section*{Acquisition of noisy information leads to decoherence effects} 

From the perspective of an external observer (an experimentalist), it is 
useful to represent the state of the system in the form of a density matrix 
${\hat\rho}=|\Psi\rangle\langle\Psi|$, where $\langle\Psi|$ is the complex 
conjugate transposition of $|\Psi\rangle$. This is because the expectation 
of a physical observable, represented by a matrix ${\hat F}$, in the state 
$|\Psi\rangle$ is given by $\langle\Psi|{\hat F}|\Psi\rangle={\rm tr}({\hat\rho}
{\hat F})$, and this is the quantity that would be measured in laboratories. 
If the observer knows the state of the system, then writing the expectation 
in terms of a pure state $|\Psi\rangle$ or in terms of a density matrix 
${\hat\rho}=|\Psi\rangle\langle\Psi|$ makes no difference. However, if the 
observer is uncertain about the state of the system, and the uncertainty 
being represented by a distribution over pure states, then we 
need to average not the wave function $|\Psi\rangle$ but the matrix 
$|\Psi\rangle\langle\Psi|$ using this distribution, and the resulting 
averaged matrix is what \textit{defines} the density matrix. 
This follows on account of the fact that observable quantities are not 
linear but quadratic in the wave function, and the square of the mean is 
not the same as the mean of the square. 

In the energy basis, the initial state 
${\hat\rho}$ thus has the matrix elements $\sqrt{p_ip_j}$, which transform 
into $\sqrt{\pi_i(\xi)\pi_j(\xi)}$ after information acquisition. 
The external observer, however, has no 
information about the value of $\xi$, so at best they may consider the 
average ${\mathbb E}[\!\sqrt{\pi_i(\xi)\pi_j(\xi)}]$ over all values $\xi$ can 
take, and this gives the matrix elements of the state 
${\hat\rho}'={\mathbb E}[|\Psi\rangle \langle\Psi|]$ as 
represented by the observer after the information acquisition by the system. 
Here, ${\mathbb E}[-]$ denotes expectation over the random variable $\xi$, 
which in the present example has the density $p(\xi)=\sum_ip_if(\xi-x_i)$. 
Then from the Bayes formula, a calculation 
shows that the transformation of the matrix elements takes the form 
$\sqrt{p_ip_j} \to{\mathbb E}[\!\sqrt{\pi_i(\xi)\pi_j(\xi)}]=
\sqrt{p_ip_j}\Lambda_{ij}$. Here, the symmetric matrix 
$\Lambda_{ij}$ is given by 
\[ 
\Lambda_{ij} = \int \! \sqrt{f(\xi)\,f(\xi+\omega_{ij})}\,  \rd \xi , 
\] 
where $\omega_{ij}=E_i-E_j$ and $f(x)$ is the density for $\epsilon$. 
Evidently, we have $\Lambda_{ii}=1$ for all $i$ and $\Lambda_{ij}<1$ for 
$i\neq j$ (assuming that $E_i\neq E_j$). 
Therefore, acquisition of information by the system will 
result in the damping of the off-diagonal elements of the density matrix. 
This is the so-called decoherence effect. In a typical (but not always the) 
case, the damping factor $\Lambda_{ij}$ of the off-diagonal element is 
more pronounced for wider energy gaps $\omega_{ij}$. 

Evidently, every 
biological process will exhibit a form of decoherence effect because 
an external observer has no access to the precise information 
acquired by the biological system, although the effect may not manifest 
in the form of a density matrix. The point is that there is a kind of 
trade-off of information flow: acquisition of information by the system 
will result in loss of information by the external observer. The amount of 
information gained by the system is the reduction of the Shannon-Wiener 
entropy associated with the transformation $p_i \to \pi_i(\xi)$, and when 
averaged over $\xi$ this is given 
by $S_\epsilon-S_\xi$, where $S_\epsilon=-\int f(x)\log f(x)\,\rd x$ is 
the entropy (measure of uncertainty) of noise and 
$S_\xi=-\int p(y)\log p(y)\,\rd y$ is the entropy of the information $\xi$. 
This follows simply by averaging the entropy change 
\[
\Delta S = -\sum_i \pi_i(\xi) \log \pi_i(\xi) + \sum_i p_i \log p_i . 
\]
In contrast, upon system's acquisition of information, the observer 
loses track of the state of the system, and hence loses information, and 
this is measured in terms of the von Neumann entropy, given by 
$-{\rm tr}({\hat\rho}\ln{\hat\rho})$. The value of the von Neumann entropy, 
however, does not agree with the change of the Shannon-Wiener entropy, 
because the density matrix ${\hat\rho}$ does not contain all the 
distributional information of the density function $p(\xi)$.

\section*{Quantum evolution is efficient in that it corresponds 
to optimal signal detection} 

More generally, for a broad range of quantum systems immersed in a general 
open environment, the state of the system as perceived by an external 
observer is described by a density matrix, whose dynamical 
evolution is described by the Lindblad equation \cite{BP}. This equation is 
deterministic, because the effect of noise is averaged over all random 
realisations of the noisy information when transforming 
from a wave function to a density matrix. Given this deterministic equation, 
finding a randomly-evolving wave function $|\Psi\rangle$ such that the 
associated density matrix ${\mathbb E}[|\Psi\rangle \langle\Psi|]$ 
obeys the Lindblad equation is known as the 
stochastic unravelling problem. In other words, can we find a stochastic 
evolution of the wave function of a system such that when the effect of 
stochasticity is averaged over, the dynamics obeys the Lindblad equation? 

It has been demonstrated recently that solutions to the stochastic 
unravelling of the Lindblad equation for open quantum dynamics are 
necessarily those for the optimal signal processing \cite{DCB}. That is 
to say, a noisy evolution of the wave function that gives rise to the Lindblad 
equation is one that corresponds to the optimal detection of the condition of 
the environment. Given that the Lindblad equation describes the dynamics 
of a very wide range of quantum systems \cite{BP}, the fact that the 
equation is underpinned by optimal signal detection supports our 
hypothesis that quantum systems evolve so as to on average maximally 
reduce the uncertainty of the conditions of the environments as encoded 
in the wave function. 

The optimality condition here is important, because it establishes the 
informational efficiency of quantum dynamics. Hence the feature of an 
efficient adaptation to environmental conditions is not unique to biological 
systems, and is embedded at the quantum-mechanical level. In 
fact, we would argue that this is why biological systems, built out of 
efficient quantum particles, are able to efficiently process information. 

\section*{From quantum to biological dynamics} 

The foregoing observations suggest that the dynamical equation 
governing the evolution of the wave function of a quantum system 
is associated with the optimal signal detection under noisy exchange 
of information with the environmental particles -- about the conditions 
of the environment. If so, then a more realistic description of a quantum 
system is obtained by the specification of the noisy information flow 
concerning the system-environment interaction, along with the 
application of optimal signal detection. This then provides a more 
complete characterisation of a quantum system, as opposed to the 
mere specification of system energy. Putting it differently, because of 
the prevalence of noise, what matters ultimately is not the energy but the  
\textit{information about the energy}, or more generally information about 
the state of the environment. If this is the case, then the adaptive 
dynamics of a system, based on noisy information about the conditions 
of the environment, is not unique to biology, but is a universal feature 
applicable equally to quantum systems as well as to biological. That is, 
signal transduction is applicable universally across all scales. This, in 
turn, leads to our proposal that mathematical models used to describe 
the dynamics of open quantum systems can be applied, 
\textit{mutatis mutandis}, to model the dynamics of biological systems. 

Motivated by these hypothesises, stochastic Schr\"odinger equations have 
been applied recently to model heliotropic and gravitropic motions of green 
plants \cite{DCB}. These models have the advantage of predicting the 
statistics of biological dynamics. Further, one can use these models to 
work out the amount of information required to be processed in order to 
achieve certain objectives (for example, for a Sunflower to identify the 
location of the sun and turn towards it). While these models are effective 
in some ways, it is legitimate to ask whether any of the quantum effects 
might play a role in the biological context. The fact that signal detection 
underpins the dynamics of quantum systems as well as those of biological 
systems, in itself, does not imply that quantum effects manifest themselves 
in the biological context. 

\section*{Manifestation of a quantum effect in biology} 

To better understand the role that quantum features might play in biology 
and ecology, we remark first that the stochastic Schr\"odinger equations 
are specified by two drivers, namely, the Hamiltonian ${\hat H}$ and the 
Lindblad operator ${\hat L}$. Now one of the fundamental features of 
quantum theory is that observable quantities need not be compatible. So 
for instance while in classical physics a particle will always have a definite 
position and a definite momentum, in quantum theory there exists 
no state (wave function) of a particle such that its position and momentum 
take definite values (i.e. with no uncertainty) simultaneously. In the case 
of a stochastic Schr\"odinger equation, we have the 
operators ${\hat H}$ and ${\hat L}$. They can either be compatible so that 
${\hat H}{\hat L}={\hat L}{\hat H}$ (the matrix product is commutative), or 
incompatible so that ${\hat H}{\hat L} \neq{\hat L}{\hat H}$ (matrices do not 
commute). In the latter case, no state can take definite values for both 
${\hat H}$ and ${\hat L}$. 

When a stochastic Schr\"odinger equation 
is applied in the biological context, the Hamiltonian ${\hat H}$ models the 
change of environmental conditions itself, while the Lindblad operator 
${\hat L}$ models adaptation in response to environmental cues. 
It is more often the case that these two effects are not compatible; 
biological systems are capable of adaptation so long as the environmental 
conditions do not change too abruptly. Otherwise, a rapid change of 
conditions can be catastrophic for survival. Hence there are two phases, 
one phase in which adaptation is feasible, and another phase in which 
adaptation is impossible. Such an effect can be modelled with a stochastic 
Schr\"odinger equation for which, and only for which ${\hat H}$ and 
${\hat L}$ are incompatible. That is, depending on the relative magnitudes 
of ${\hat H}$ and ${\hat L}$, the dynamics of the system exhibits different 
characteristics. Although detailed features of such phase transitions have 
not been fully uncovered, their existence is well documented \cite{Bassi,BL}. 

Therefore, while from a purely signal-detection perspective there is little 
evidence that some of the counter-intuitive quantum features (such as 
entanglement or phase coherence) play any role in biology and ecology, 
the lack of compatibility among observable quantities -- another quantum 
feature -- does appear to play an important role. This observation makes 
the stochastic Schr\"odinger equation a useful and effective mathematical 
tool for characterising biological dynamics, even though in many cases 
the dynamics can equally be modelled using the language of classical 
signal detection. 

\section*{Quantifying information contained in environmental cues} 

Whether classical or quantum signalling models are used to characterise 
biological dynamics, the use of these models has the advantage -- apart 
from being able to predict the statistics of future events -- of allowing 
for the quantification of information. This is of particular interest in the case 
of plants, for, we do not normally associate the notion of `information 
processing' to plants, and yet, plants do process environmental information 
at all times. Thus, models can be used to assist estimating the 
information contents of environmental cues. Take, for instance, the example 
considered earlier in which the signal is modelled by the random variable 
$X$, taking the value $x_i$ with the probability $p_i$, and noise by 
$\epsilon$, with the density $f(x)$. Then the information contained in the 
noisy 
observation $\xi=X+\epsilon$ about the value of $X$ is measured by the 
mutual information $J(\xi,X)$. This is the amount of information contained 
in environmental cues concerning the signal of interest for the plant. 
Writing $p(\xi)=\sum_i p_i f(\xi-x_i)$ for the 
density of $\xi$, the mutual information can be shown to be given by the 
entropy difference \cite{GY,BDFH} 
\[ 
J(\xi,X) = \int f(x)\log f(x) \, \rd x - \int p(\xi) \log p(\xi) \, \rd \xi . 
\] 
Remarkably, this agrees with the negative of the average change of 
entropy obtained earlier. 
Therefore, once a model for the signal and noise is chosen, one can 
predict the statistics of biological dynamics by using the \textit{a posteriori} 
probabilities. Model parameters can then be calibrated by comparing the 
statistics thus obtained against empirical data. When a model is calibrated, 
it can be used to quantify the information content. In other words, even 
though there is neither an intentional transmitter of the signal nor a 
purposefully built communication channel, it is nevertheless possible to 
estimate the information content of environmental cues. 

\section*{Discussion} 

The routine threats experienced by developing plants are numerous. 
Compared to animals, the sessile plant only requires the physical 
constituents of their environments: light, ${\rm CO}_2$, 
${\rm H}_2{\rm O}$ and minerals \cite{Gilroy}.  
A serious lack or imbalance of any of these can threaten viability as can 
a substantial number of different mechanical, thermal, and biological 
challenges. 
Phenotypic plasticity is a common behavioural response; 
tune the phenotype to the experienced environmental conditions. 
Alterations occur 
either via numbers of branch roots, shoot branches, leaves or buds, or 
by changing their physical parameters, such as leaf area, stem thickness, 
stiffness or proportions of different cell types. For instance, a lack of 
water often results in increased root proliferation which penetrate 
to increased soil depths. 
Plants are the basis of all food chains so predation is common and often 
damaging. When predatory damage is recognised many plants synthesise 
natural pesticides such as capsaicin, nicotine, or caffeine in response. These 
kill the damaging insect or deter it from additional feeding.  It is estimated 
that as many as 100,000 such natural pesticides exist, the product of an 
arms race between plants and predatory insects for 300 million years 
\cite{Ames}. 
Volatiles are also synthesised, and these attract predators of the pest. In 
essence they function like a burglar alarm. Summing up, the capability of 
efficiently processing information is vital to the survival of plants. 

The information-theoretic approach to understanding biological dynamics 
is useful because it offers a way of modelling the dynamical behaviours of 
biological systems and enables predictions. We believe it is particularly 
advantageous in shedding light on the behaviour of green plants, for, it is 
otherwise difficult to assess or measure informational quantities associated 
with them. In particular, the mutual information involves the entropy of the 
noisy observation $\xi$, which is certainly difficult to estimate from data. 
But what do we learn from these information measures about plant 
biology? And what are the open challenges from the informational 
perspective? 

In view of the fact that plants form the basis of almost all food chains, 
enhanced understandings of their behaviours, efficiencies, and capabilities 
are becoming increasingly more important \cite{Gilroy}. 
This information-based approach will naturally contribute towards this endeavour. In particular, as well as processing information, we expect that biological systems erase some of the processed information. This is to be expected with transduction systems using $[{\rm Ca}^{2+}]_{\rm i}$ since at the end of the first transient the system is primed to respond to alternative signals.  But erasure of information inevitably requires energy consumption and heat production, in accordance with Landauer’s principle in computation \cite{LR}. In particular, the less efficient is the computational capacity, the more energy it requires. Based on the recent experiments with bean plants that can identify an object in its vicinity to climb up \cite{Calvo}, 
it has been conjectured that green plants must be able to process information more efficiently than, say, the best computers available \cite{DCB}. Beyond this one estimate, little has been investigated or explored in this direction. However, the information-based approach offers a new avenue of research that can be supplemented by theoretical models \cite{Fields}. 

Determining the computational efficiency of green plants is an important component in understanding how they can adapt to changing environments. The observation made here that techniques of communication theory can be applied to model biological dynamics (whether classical or quantum models are used), even when there are no active agents transmitting the signal or purposefully built communication channels, opens up the possibility of making a significant progress in investigating and determining the informational efficiencies, capabilities, and limits of plants. A deeper understanding of these issues may even uncover ideas for technological advances. To make a significant progress in this area, however, more data are required on plants. These include, in particular, the understanding of how information is processed, how it is stored, and how they are erased, in plants. Equally valuable information concerns the number of information-storing units in plants (for processed information, not genetic), and more generally how much energy is consumed by plant cells. For instance, with a nanokelvin-resolution thermometer \cite{Reihani} 
it may be possible to measure heat production associated with erasure of information in plants, which in turn will elucidate the computational efficiencies of plants, or any other biological system for that matter. We hope to explore these ideas elsewhere.

\vspace{0.5cm}
\begin{footnotesize}
\noindent {\bf Acknowledgements}. The authors thank Jim Al-Khalili, 
Bernhard Meister, and Simon Saunders for stimulating discussions. 
DCB acknowledges support from the John Templeton Foundation 
(grant 62210). The opinions expressed in this publication are those of 
the authors and do not necessarily reflect the views of the 
John Templeton Foundation.
\end{footnotesize}

\end{document}